\documentclass[aps,12pt,superscriptaddress]{revtex4}

\usepackage{graphics}
\usepackage{graphicx}
\usepackage{mathalpha,amsmath,amssymb}
\usepackage[a4paper, total={6in, 8.37in}]{geometry}
\usepackage{xcolor,soul}
\usepackage{hyperref}

\begin{document}

\title{Double $\mathbf{k}$-Grid Method for Solving the Bethe-Salpeter Equation via Lanczos Approaches} 

\author{Ignacio M. Alliati}
\affiliation{School of Mathematics and Physics, Queen’s University Belfast, Belfast BT7 1NN, Northern Ireland, United Kingdom}
\author{Davide Sangalli}
\altaffiliation[Also at ]{European Theoretical Spectroscopy Facility (ETSF)}
\affiliation{Istituto di Struttura della Materia—Consiglio Nazionale delle Ricerche (CNR-ISM), Division of Ultrafast Processes in Materials (FLASHit), Via Salaria Km 29.5, CP 10, I-00016 Monterotondo Stazione, Italy}
\author{Myrta Gr\"uning}
\altaffiliation[Also at ]{European Theoretical Spectroscopy Facility (ETSF)}
\affiliation{School of Mathematics and Physics, Queen’s University Belfast, Belfast BT7 1NN, Northern Ireland, United Kingdom}

\begin{abstract}
Convergence with respect to the size of the k-points sampling-grid of the Brillouin zone is the main bottleneck in the calculation of optical spectra of periodic crystals via the Bethe-Salpeter equation (BSE). We tackle this challenge by proposing a double grid approach to k-sampling compatible with the effective Lanczos-based Haydock iterative solution. Our method relies on a coarse k-grid that drives the computational cost, while a dense k-grid is responsible for capturing excitonic effects, albeit in an approximated way. Importantly, the fine k-grid requires minimal extra computation due to the simplicity of our approach, which also makes the latter straightforward to implement. We performed tests on bulk Si, bulk GaAs and monolayer MoS$_2$, all of which produced spectra in good agreement with data reported elsewhere. This framework has the potential of enabling the calculation of optical spectra in semiconducting systems where the efficiency of the Haydock scheme alone is not enough to achieve a computationally tractable solution of the BSE, e.g., large-scale systems with very stringent k-sampling requirements for achieving convergence.
\end{abstract}
\maketitle

\section{Introduction}\label{sc:intro}
Many-body perturbation theory (MBPT) offers the right framework for treating neutral excitations via Green's function methods\cite{martin2016interacting,Onida2002,Reining2018,Golze2019,Marini2009}. This requires solving the Bethe-Salpeter equation (BSE) \cite{Salpeter1951,Hedin1965,Hedin1971}, which relies on a two-particle propagator to account for the presence of electron-hole pairs (i.e., excitons). The description of excitonic effects is crucial to compute optical spectra in extended systems, particularly in semi-conductors and insulators, for which methods based on the Random Phase Approximation (RPA) or time-dependent (TD-) density functional theory (DFT) with (semi-)local exchange-correlation functionals tend not to agree with experimental results \cite{martin2016interacting,Onida2002}. Calculations within the BSE framework are generally much more cumbersome and computationally demanding than DFT ones, and it is rather easy to reach the limits of what can be practically computed. Hence, the need for convergence studies is a key aspect of every MBPT calculation to alleviate the computational burden as much as possible while, at the same time, trying to ensure an accurate description of the system at hand. In general, electronic structure calculations in periodic systems treated plane waves require convergence with respect to the size of this basis set as well as the sampling of the Brillouin zone (BZ). In particular, MBPT methods also require the inclusion of unoccupied states in the form of an, in principle, infinite summation that needs to be truncated to the minimum value that nonetheless captures the physics at play.
Furthermore, the solution of the BSE requires far denser $\mathbf{k}$-sampling than DFT calculations to achieve an accurate description of excitons. This is because the excitonic wave-functions are usually quite spread out, with a periodicity well beyond the unit cell, and in order to expand them in a basis of transitions $\{vc\mathbf{k}\}$ (electron-hole space), very dense $\mathbf{k}$-grids are required. Moreover, BSE methods do not exploit symmetry so they are solved in the full BZ. The cubic scaling of the number of $\bf k$-points in bulk systems from one grid to the next makes matters even worse (the quadratic scaling in 2D systems is more manageable). Such $\bf k$-grid requirements may still be feasible for small systems, with few atoms per unit cell and few valence electrons per atoms. However, medium to large size unit cells of atoms with many valence electrons (e.g. transition metals) become prohibitively costly as the number of $\bf k$-points increases, and the solution of the BSE in a very dense grid (e.g. $60\times60\times60$) is simply out of reach. For all these reasons, the issue of $\bf k$-point convergence is critical for the solution of the BSE and represents the bottleneck in its computational implementation. Therefore, the introduction of alternative numerical methods and approximations that can effectively deal with $\bf k$-point convergence in the BSE is of utmost importance. 

Albeit with the limitations described above, there are currently several approaches to solve the BSE. These are, in order of decreasing computational cost, inversion, full diagonalisation and Lanczos approaches, and will be described below. A first distinction would be based on whether the equation is solved in its Dyson-like form or re-cast as a two-particle Hamiltonian in transition space. The first approach requires the inversion of the BSE kernel matrix which, depending on the size of the matrix, can become impracticable. In such cases one would turn to the Hamiltonian formulation of the problem (see, for example, ref. \cite{Onida2002}). In the latter, the two-particle Hamiltonian is diagonalised to obtain the eigen-values (excitonic energies) and eigen-vectors (excitonic wave-functions). If the BSE matrix of a given system is still too big for full diagonalisation, one can resort to Lanczos \cite{Lanczos1950} approaches which are usually a cost-effective option for sparse matrices \cite{cini2007}. These algorithms have been widely used for the calculation of response functions, both at the TD-DFT \cite{Rocca2008,Ge2014} and BSE \cite{Rocca2012} levels. In the latter, Lanczos approaches eliminate the need for inverting the BSE kernel or fully diagonalising the two-particle Hamiltonian. Rather, the latter is re-expressed as a tri-diagonal matrix based on recursive relations, which leads to an iterative solution of the problem that is computationally cheaper than full diagonalisation. Unfortunately, while previously described solvers produce the full set of both excitonic energies and wavefunctions of the system at hand, Lanczos schemes lead to a partial solution of the problem. For instance, Haydock's implementation \cite{Haydock1980} of the Lanczos approach provides only the full set of the eigen-values of the two-particle Hamiltonian (i.e., one obtains the full spectrum but not the excitonic wave-functions). Despite the numerical advantages of Lanczos solvers, a given system could still be too big for computing optical spectra. As the diagonalisation itself ceases to be a problem with Lanczos schemes, the bottleneck now shifts to the previous step of computing and storing the BSE kernel, which can render the calculation impracticable depending on the size of the electron-hole (e-h) basis. Nothing too extreme would be required to reach this condition, e.g., a magnetic system with around 100 electrons per unit cell, slow convergence with respect to bands and a $6\times6\times6$ \textbf{k}-grid would certainly be beyond reach. At this point, there is little alternative for solving the BSE and computing optical spectra, which is the challenge we intend to tackle in this manuscript. 

The work presented here concentrates on improving the convergence of optical response spectra calculations within the BSE with respect to the number of $\mathbf{k}$-points. This issue has been the target of many research efforts over the years. Rohlfing \textit{et al.} introduced a scheme to interpolate the BSE matrix in the BZ \cite{Rohlfing1998}. Their strategy is based on a double grid approach by which the kernel matrix elements are properly calculated on a coarse $\mathbf{k}$-grid and approximated on a fine $\mathbf{k}$-grid. As a function of $\mathbf{q}$, the $\mathbf{k}$-point difference between two transitions in e-h space, the BSE kernel is sharply peaked at the origin and a regular interpolation in the BZ would fail. However, expressing these matrix elements as $a\mathbf{q}^{-2}+b\mathbf{q}^{-1}+c$ results in the coefficients varying slowly in the BZ. These coefficients are then interpolated by virtue of knowing them exactly in the coarse $\mathbf{k}$-grid. Their approximation also considers the varying phases of the single-particle states in the BZ, which requires knowledge of the wavefunctions in the fine $\mathbf{k}$-grid. This crucial point becomes a drawback when one is limited by memory and disk storage rather than computation, which is increasingly the case nowadays. More recently, Fuchs \textit{et al.} proposed the use of hybrid $\mathbf{k}$-meshes in the form of a coarse $\mathbf{k}$-grid for the whole BZ and a denser $\mathbf{k}$-grid around the $\Gamma$-point \cite{Fuchs2008} only. Even though the kernel matrix elements are properly calculated on both grids, this method allows to refine $\mathbf{k}$-sampling only where is needed, resulting in fewer $\mathbf{k}$-points in total. The downside of using non-uniform grids becomes apparent in the calculation of the electron-hole attraction term of the BSE kernel, as knowledge of the screening at $\mathbf{q}$-points not included in the hybrid grid itself will be needed. This complication requires additional computation (or at least an interpolation) if one intends to use the RPA screening, as is the case in this work. Kammerlander \textit{et al.} applied double grid techniques to solving the BSE by inversion \cite{Kammerlander2012}. In the latter, the BSE is solved on the coarse $\mathbf{k}$-grid while the fine $\mathbf{k}$-grid us used compute the independent particle part of the two-particle response function. This technique, which also benefits from Wannier interpolation of the Kohn-Sham (KS) orbitals, has proven successful in accurately reproducing the spectra of several materials. However, as it ultimately relies on matrix inversion, its application is limited to \textit{small} systems, i.e., systems which could be computed by the inversion solver in the coarse grid, albeit underconverged. Finally, an interesting generalisation of the method in ref. \cite{Rohlfing1998} has been proposed by Gillet \textit{et al.}\cite{Gillet2016}, where the interpolation of the BSE kernel matrix element at a given fine-grid $\mathbf{k}$-point considers eight coarse-grid $\mathbf{k}$-points around it. Importantly, this method is compatible with Haydock's solution scheme to the BSE. Moreover, substantial savings in memory requirements and disk storage are achieved by interpolating kernel matrix elements on the fly. Nevertheless, this method still requires knowledge of the KS orbitals in the fine grid. Depending on the number of bands and density of the fine grid, this can entail prohibitive memory requirements. 

In this work, we also propose a double grid approach, including a coarse $\mathbf{k}$-grid where the BSE kernel is properly calculated and a fine, denser $\mathbf{k}$-grid where the corresponding matrix elements are approximated. At variance with ref. \cite{Kammerlander2012}, we propose an approximation that is compatible with the computationally cheapest solution to the BSE, namely Lanczos-based iterative solvers. Crucially, this allows us to target materials for which optical spectra cannot be currently computed, i.e., relatively big systems that can only be solved by Lanczos approaches and in $\mathbf{k}$-grids that fall short of a converged solution. Another distinctive feature of the method presented here is its simplicity. It is far easier to implement than previous attempts\cite{Rohlfing1998,Fuchs2008}. Importantly, the introduction of the fine $\mathbf{k}$-grid requires minimal extra computation and memory with respect to the coarse $\mathbf{k}$-grid. In particular, knowledge of the wavefunctions in the fine $\mathbf{k}$-grid is not needed, nor is the calculation of the RPA screening in any extra $\mathbf{k}$-point. Therefore, the computational cost remains roughly at the level of the coarse $\mathbf{k}$-grid, while an approximate description of broad excitons is achieved by virtue of adding a fine $\mathbf{k}$-grid. The remainder of the manuscript is structured as follows. Section~\ref{sc:methods} describes the proposed double grid approach in detail while Section~\ref{sc:results} reports the results obtained for a variety of semiconductors. Section~\ref{sc:discussion} presents the gains in computational cost that our implementation achieves. It also outlines a comparative assessment of particular choices made within the method and discusses the limitations of the approach.

\section{Methods}\label{sc:methods}
\subsection{Haydock solution of the BSE}
Optical absorption spectra are represented by the imaginary part of the macroscopic dielectric function $\mathfrak{Im}[\epsilon_{\mathrm{M}}]$, which is obtained by taking the long wavelength limit of an expression involving the microscopic dielectric function $\epsilon(\mathbf{q},\omega)$---where $\mathbf{q}$ represents the transferred momenta while $\omega$ is the frequency. For neutral excitations, $\epsilon(\mathbf{q},\omega)$ is defined in terms of the polarisation or density-density response function $\chi$, which is in turn calculated within the RPA, i.e., as a Dyson-like equation being the non-interacting polarisation $\chi^0$ a product of non-interacting one-particle Green's functions that describe the propagation of one electron or one hole. However, optical spectra of extended systems require the inclusion of excitonic effects, which will ultimately lead us to a two-particle Green's function that describes the dynamics of an electron-hole pair $\{vc\mathbf{k}
\}$ (we only consider vertical transitions at point $\mathbf{k}$ in the BZ between an occupied band $v$ and an empty band $c$). This is achieved by defining the macroscopic dielectric function via an interacting polarisation $\bar \chi$, i.e., $\epsilon_{\mathrm{M}}(\mathbf{q},\omega) \equiv 1 - v(\mathbf{q}) \bar \chi_{\mathbf{G}=0,\mathbf{G'}=0}(\mathbf{q},\omega)$. This interacting polarisation is obtained in terms of an electron-hole (e-h) Green's function $\bar L$ as in Eq. \ref{eq:ec_intpol},
\begin{equation}
  \lim_{\mathbf{q}\to 0} \bar\chi_{\mathbf{G}=0,\mathbf{G'}=0}(\mathbf{q},\omega) = 
- i \sum_{nm\mathbf{k}}\sum_{n'm'\mathbf{k'}} \lim_{\mathbf{q}\to 0} [\rho^*_{nm\mathbf{k}}(\mathbf{q},\mathbf{G}=0) \rho_{n'm'\mathbf{k'}}(\mathbf{q},\mathbf{G'}=0] \bar L_{\substack{ nm\mathbf{k}\\n'm'\mathbf{k'}}}(\omega) .
  \label{eq:ec_intpol}
\end{equation}
In Eq. \ref{eq:ec_intpol}, $\rho_{nm\mathbf{k}} (\mathbf{q,G}) = \langle n\mathbf{k}| e^{i\mathbf{(q+G)\cdot r}}|m\mathbf{k-q} \rangle$ are the oscillator strengths. For simplicity, unpolarised electrons are assumed in the discussion, however we stress that the method is not limited to non-magnetic systems. The Bethe-Salpeter equation is then the Dyson-like equation for $\bar L$, 
\begin{equation}
  \bar L_{\substack{ nm\mathbf{k}\\n'm'\mathbf{k'}}}(\omega) = 
  L^0_{nm\mathbf{k}}(\omega) [\delta_{nn'}\delta_{mm'}\delta_{\mathbf{k}\mathbf{k'}} + i \sum_{vc\mathbf{k_1}} \Xi_{\substack{ nm\mathbf{k}\\vc\mathbf{k_1}}}(\omega) \bar L_{\substack{ vc\mathbf{k_1}\\n'm'\mathbf{k'}}}(\omega)],
  \label{eq:ec_BSE}
\end{equation} 
where the matrix $\Xi$ is the so called BSE kernel,
 \begin{equation}
    \Xi_{\substack{ nm\mathbf{k}\\vc\mathbf{k_1}}} = W_{\substack{ nm\mathbf{k}\\vc\mathbf{k_1}}}
    -2\bar V_{\substack{ nm\mathbf{k}\\vc\mathbf{k_1}}},
  \label{eq:ec_BSEkernel}
 \end{equation}
\begin{equation}
    W_{\substack{ nm\mathbf{k}\\vc\mathbf{k_1}}} = \frac{1}{\Omega N_q}\sum_{\mathbf{G,G'}} \rho_{nv\mathbf{k}}(\mathbf{q=k-k_1,G})
    \rho^*_{mc\mathbf{k_1}}(\mathbf{q=k-k_1,G'})
    \epsilon^{-1}_{\mathbf{G,G'}}v(\mathbf{q+G'}),
    \label{eq:ec_W}
\end{equation}

\begin{equation}
    \bar V_{\substack{ nm\mathbf{k}\\vc\mathbf{k_1}}} = \frac{1}{\Omega N_q}\sum_{\mathbf{G\neq 0}} \rho_{nm\mathbf{k}}(\mathbf{q=}0,\mathbf{G})
    \rho^*_{vc\mathbf{k_1}}(\mathbf{q=}0,\mathbf{G})
    v(\mathbf{G}).
    \label{eq:ec_V}  
\end{equation}
The BSE kernel is written as shown in Eq. \ref{eq:ec_BSEkernel} and its two contributions, namely the e-h attraction $W$ and the e-h exchange $\bar V$, can be calculated as in Eqs. \ref{eq:ec_W} and \ref{eq:ec_V}.The solution of this Dyson-like equation would require to invert the BSE kernel, which can be prohibitively costly as explained in Section~\ref{sc:intro}. Hence, the problem is re-cast in terms of a two-particle Hamiltonian in e-h space,
\begin{equation}
    H^{2p}_{\substack{ nm\mathbf{k}\\n'm'\mathbf{k'}}}
    = 
    E_{nm\mathbf{k}} \; \delta_{nn'}\delta_{mm'}\delta_{\mathbf{kk'}}
    +\; (f_{n\mathbf{k}} - f_{m\mathbf{k}}) \; \Xi_{\substack{ nm\mathbf{k}\\n'm'\mathbf{k'}}},
    \label{eq:ec_2pH}  
\end{equation}
where $E_{nm\mathbf{k}}$ is the energy of the vertical transition from band $n$ to band $m$ at point $\mathbf{k}$ according to either the KS or quasi-particle (QP) energies.

Diagonalising this matrix would provide the excitonic eigen-values and eigen-vectors required to compute optical spectra, however in this work, we focus on Lanczos-based methods. In particular, Haydock's algorithm \cite{Haydock1980,Benedict1998,Benedict1999} consists on an iterative method based on a set of recursive relations, namely,
\begin{equation}
    a_n=\langle V_n|H^{2p}|V_n\rangle ,
    \label{eq:ec_recursiverelations1}
\end{equation}
\begin{equation}
    b_{n+1}=\|(H^{2p}-a_n)|V_n\rangle -b_n |V_{n-1}\rangle \| ,
    \label{eq:ec_recursiverelations2}
\end{equation}
\begin{equation}
    |V_{n+1}\rangle = \frac{1}{b_{n+1}}[(H^{2p}-a_n)|V_n\rangle -b_n |V_{n-1}\rangle],
    \label{eq:ec_recursiverelations3}
\end{equation}
with $n$ being the iteration index. This set of equations corresponds to Hermitian Hamiltonians (the pseudo-Hermitian case has a slightly more complicated form \cite{Gruning2011}). Eqs. \ref{eq:ec_recursiverelations1}-\ref{eq:ec_recursiverelations3} allow for the calculation of the factors $a$ and $b$, and the so called Haydock vector for the next iteration $|V_{n+1}\rangle$. The initial Haydock vector is calculated as $|V_0\rangle = \frac{|P\rangle}{\|P\|}$ being $|P\rangle$ the vector defined as
\begin{equation}
    |P\rangle = \sum_{vc\mathbf{k}} \lim_{\mathbf{q}\to 0} \frac{1}{|\mathbf{q}|} \rho^*_{vc\mathbf{k}}(\mathbf{q,G}=0) |vc\mathbf{k}\rangle.
    \label{eq:ec_HaydockP}
\end{equation}
On each iteration $n$, the optical spectrum is calculated according to,
\begin{equation}
    \epsilon^{(n)}_M(\omega)=1-\|P\|^2 \frac{1}{(\omega -a_1)-\frac{b^2_2}{(\omega-a_2)-\frac{b_3^2}{...}}},
    \label{eq:ec_HaydockSpectrum}
\end{equation}
until the difference between spectra of successive iterations is below an acceptable threshold.

\subsection{Double grid approach}\label{sc:dgrid}
First, we consider a coarse $\mathbf{k}$-grid where no approximations are applied, i.e., the BSE kernel is computed for all vertical transitions involving $\mathbf{k}$-points in this grid, which requires knowledge of the KS orbitals and energies (potentially corrected to QP energies) at each of these $\mathbf{k}$-points. The solution of the BSE in this grid would typically be computationally manageable but produce underconverged optical spectra. Thus, a much denser fine $\mathbf{k}$-grid will be added to the system. We will denote $\mathbf{k}$-points belonging to the fine grid with the letter $\mathbf{\kappa}$, while those in the coarse grid will be labelled $\mathbf{K}$. Moreover, $\mathbf{\kappa}$-points will be grouped in domains centred around the $\mathbf{K}$-points in such way that $\mathbf{Dom}(\mathbf{K_i})$ will be composed by the $\mathbf{\kappa}$-points that are closer to $\mathbf{K_i}$ than to any other $\mathbf{K}$-point. The number of  $\mathbf{k}$-points in this fine grid would ordinarily be too large for the BSE to be solved in full, and hence, approximations will be introduced for the fine grid. The two-particle Hamiltonian in Eq. \ref{eq:ec_2pH} can be thought of as a shift (the diagonal matrix containing the energies of each transition) plus a rotation (the BSE kernel). The approximation proposed implies that the diagonal matrix is calculated in the fine grid, for which knowledge of the KS energies of each band at every $\mathbf{\kappa}$-point in the fine grid is required. The BSE kernel, however, will not be calculated in full but rather, every matrix element involving at least one transition in the fine grid will be approximated according to some rules for kernel extension. This allows the method to dispense with the KS orbitals in the fine grid, which has a great impact on memory requirements.

The way in which the BSE kernel is extended from the coarse to the fine grid has been carefully considered as it has significant impact on the results. The best agreement with experimental spectra was achieved with an approach we refer to as diagonal kernel extension (DKE). Let us consider one $\mathbf{k}$-point in the coarse grid, $\mathbf{K_I}$. There will be a group of $\mathbf{\kappa}$-points in the fine grid that map to it, namely those in the domain $\mathbf{Dom}(\mathbf{K_I})$. We will label those with a second numerical sub-index as $\kappa_{\mathbf{I}_1},\kappa_{\mathbf{I}_2},\kappa_{\mathbf{I}_3},...,\kappa_{\mathbf{I}_i},...$. Given that the fine grid contains the coarse grid, we have that $\kappa_{\mathbf{I}_1}=\mathbf{K_I}$, while $\kappa_{\mathbf{I}_i}$ with $i \neq 1$ are other fine grid points close to $\mathbf{K_I}$. Having established the nomenclature in this way, then DKE would imply the definition
\begin{equation}
    \Xi_{\substack{ nm\kappa_{\mathbf{I}_i}\\n'm'\kappa_{\mathbf{I'}_{i'}}}} \equiv \;
    \Xi_{\substack{ nm\mathbf{K_I}\\n'm'\mathbf{K_{I'}}}} \; \delta_{ii'},
    \label{eq:ec_DKE}
\end{equation}
where the R.H.S is known and calculated exactly while the L.H.S is the unknown matrix element we are trying to approximate (see SI for a visual representation of Eq. \ref{eq:ec_DKE}). Thus, Eq. \ref{eq:ec_DKE} is only exact for the $\mathbf{k}$-point that belongs both the coarse and the fine grids ($i=i'=1$), and approximated otherwise. Even though the BSE kernel is not, in general, a diagonally-dominant matrix, it is true that the diagonal matrix elements usually have values orders of magnitude higher than those of immediately close off-diagonal elements. The DKE approach preserves this character when extending the kernel from the coarse grid to the fine grid. Essentially, each matrix element of the coarse grid BSE kernel expands into a block in the fine grid matrix. The DKE method ensures that each block is strictly diagonal, which is very relevant when expanding one of the diagonal matrix elements of the coarse grid matrix. In Section~\ref{sc:an_DKE}, the DKE is compared with a possible alternative kernel extension. 

Finally, let us discuss how this double grid method fits within the Haydock algorithm. It is apparent from Eqs. \ref{eq:ec_recursiverelations1}-\ref{eq:ec_recursiverelations3} that this scheme relies mainly on the matrix vector multiplication $H^{2p} |V_n\rangle$, so we will focus on how this is adapted to account for the fine grid. The two-particle Hamiltonian has already been described above, i.e., the BSE kernel is approximated by DKE (Eq. \ref{eq:ec_DKE}) and the diagonal part needs no approximation as the KS (or QP) energies are known in the fine grid. All there is left is then to define how the Haydock vectors $|V_n\rangle$ are extended to the fine grid and initialised. The initial Haydock vector $|V_0\rangle$ is calculated in the coarse grid according to Eq. \ref{eq:ec_HaydockP}. Each component is associated to one transition $vc\mathbf{k}$ and thus, when moving from the coarse to the fine grid, the number of components will increase according to the ratio between the number of $\mathbf{\kappa}$-points and $\mathbf{K}$-points. From Eq. \ref{eq:ec_HaydockP}, it is clear that the KS orbitals at the $\mathbf{\kappa}$-points would be required to properly initialise the Haydock vector in the fine grid. As these orbitals are not available in our method, those components will be initialised as being equal to the corresponding transition in the coarse grid. In other words,
\begin{equation}
    |P\rangle_{\text{FG}} = \sum_{vc\mathbf{K_I}}  \lim_{\mathbf{q}\to 0} \frac{1}{|\mathbf{q}|} \rho^*_{vc\mathbf{K_I}}(\mathbf{q,G}=0)
    \sum_{\substack{\kappa_{\mathbf{I}_i} \; \in \\ \mathbf{Dom}(\mathbf{K_I})}} |vc\kappa_{\mathbf{I}_i}\rangle,
    \label{eq:ec_initialisation_FG}
\end{equation}
where FG denotes the fine grid.
It is apparent that $|P\rangle_{\text{FG}}$ has many more components than $|P\rangle$, due to each coarse grid transition (at $\mathbf{K_I}$) being replicated into many transitions at all the $\mathbf{\kappa}$-points in the domain of $\mathbf{K_I}$. The recursive relations in Eqs. \ref{eq:ec_recursiverelations1}-\ref{eq:ec_recursiverelations3} would formally require the multiplication of the fine grid (full) BSE kernel times Haydock vectors of the size of $|P\rangle_{\text{FG}}$. 
In our implementation, we calculate the matrix-vector multiplication without storing  the BSE kernel on the fine grid. 
Let us divide a given Haydock vector $|V\rangle$ in \textit{fragments} according to the $\mathbf{\kappa}$-point of each transition. Formally, this would mean projecting $|V\rangle$ over the different transitions (i.e, the abstract vectors $|vc\mathbf{\kappa_{I_i}}\rangle$ that form the new/extended basis of e-h space) and then grouping components by their $\mathbf{\kappa}$-point as \textit{fragments}. These fragments are convenient for the matrix vector multiplication. In fact, this operation can be expressed as
\begin{equation}
    r_{nm\kappa_{\mathbf{I}_{i}}} =
    \sum_{n'm'\kappa_{\mathbf{I'}_{i'}}} \Xi_{\substack{ nm\kappa_{\mathbf{I}_i}\\n'm'\kappa_{\mathbf{I'}_{i'}}}} c_{n'm'\kappa_{\mathbf{I'}_{i'}}} ,
    \label{eq:ec_KbyV1}
\end{equation}
where $c_{vc\mathbf{\kappa_{I_i}}}=\langle vc\mathbf{\kappa_{I_i}}|V\rangle$ are the components of the vector to be multiplied and $r$ are, analogously, the coefficients of the resulting vector.
Applying the DKE to the BSE matrix (Eq. \ref{eq:ec_DKE}), we obtain 
\begin{equation}
    r_{nm\kappa_{\mathbf{I}_{i}}} =
    \sum_{n'm'\mathbf{K}_{\mathbf{I'}}} 
    \sum_{\substack{i' \; \in \\ \mathbf{Dom}(\mathbf{K_{I'}})}}
    \Xi_{\substack{ nm\mathbf{K}_{\mathbf{I}}\\n'm'\mathbf{K}_{\mathbf{I'}}}} \delta_{i,i'} \; c_{n'm'\kappa_{\mathbf{I'}_{i'}}} =
    \sum_{n'm'\mathbf{K}_{\mathbf{I'}}} 
    \Xi_{\substack{ nm\mathbf{K}_{\mathbf{I}}\\n'm'\mathbf{K}_{\mathbf{I'}}}} c_{n'm'\kappa_{\mathbf{I'}_{i}}},
    \label{eq:ec_KbyV2}
\end{equation}
where matrix elements in the R.H.S are those of the coarse-grid BSE kernel and the resulting summation runs over the $\mathbf{K}$-points in the coarse grid only. Computationally, this means adding a loop over the $\mathbf{\kappa}$-points in the domain of each $\mathbf{K}$.

\section{Results}\label{sc:results}
The double grid method proposed here to calculate optical spectra via the BSE has been implemented in the Haydock solver of the Yambo code \cite{Marini2009,Sangalli2019} and tested on a variety of semiconductors. In this section, we present the resulting optical spectra of each material. We note that the spectra produced by this approach represented a sharp improvement with respect to the coarse grid solution, while requiring only a marginal increase in computational cost. Although Gamma-centred $\mathbf{k}$-grids were used throughout this study, our method can also be used with shifted grids (see an example in SI).
\subsection{Si bulk}
\begin{figure}[h]
    \centering
    \includegraphics{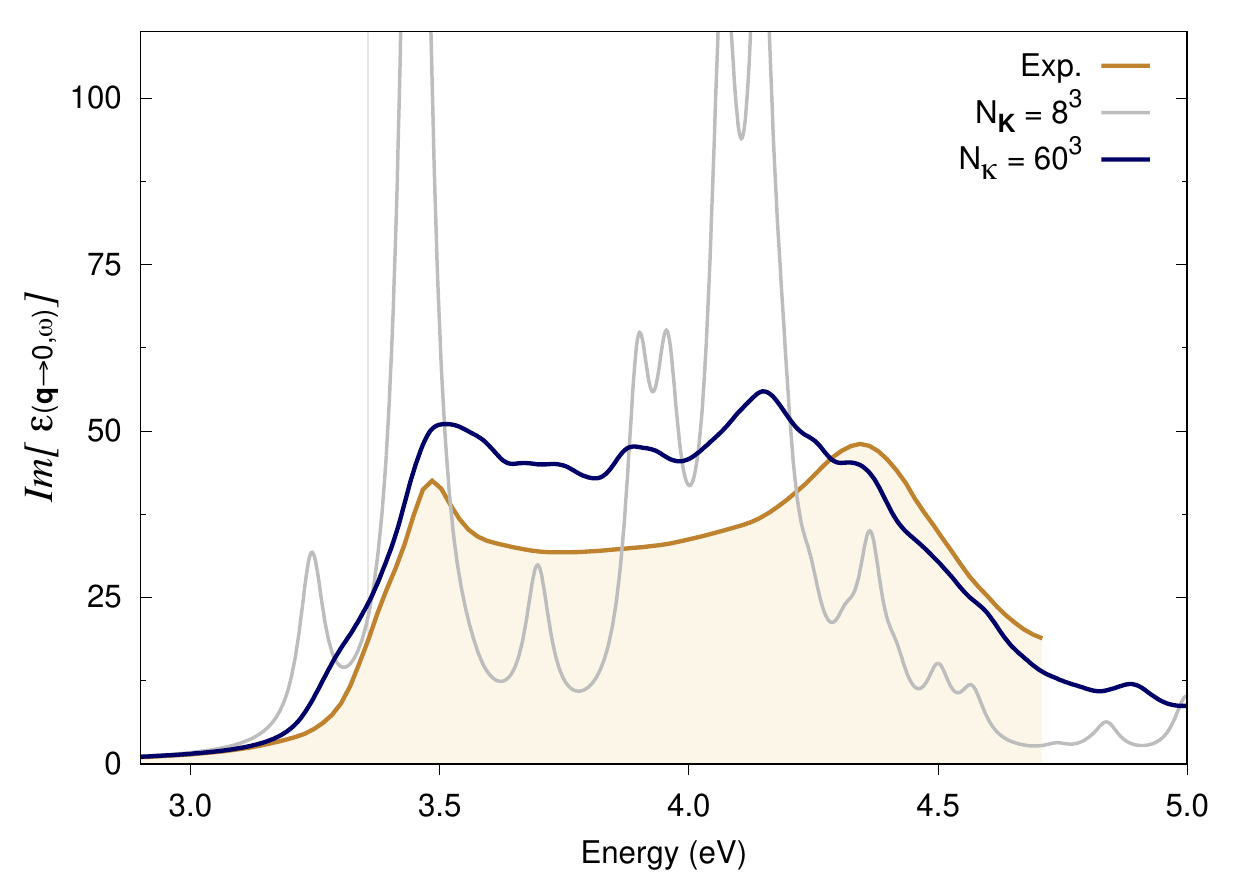}
    \caption{Optical absorption spectra of bulk Si. BSE spectra calculated using the double-grid approach described in this work are assessed against the experimental spectrum at 10 K from Ref.~\cite{Jellison1983}. Spectra are calculated on an $8\times 8\times 8$ coarse $\mathbf{k}$-grid and a  denser fine $\mathbf{k}$-grid, indicated by $N_\mathbf{\kappa}$. $N_\mathbf{K}=8^3$ corresponds to a standard calculation on an $8\times 8\times 8$ $\mathbf{k}$-grid while $N_\mathbf{\kappa}=60^3$ corresponds to a double-grid calculation with a $60\times 60 \times 60$ fine grid. We consider all $e-h$ pairs from the top 4 valence bands to the 4 bottom conduction bands. All $\mathbf{k}$-grids are Gamma-centred.}   
    \label{fig:1silicon}
\end{figure}

It is notoriously difficult to converge the optical spectrum of bulk Si with respect to $\mathbf{k}$-points since a very dense $\mathbf{k}$-sampling is required to properly describe its excitons. The starting point for our Si calculations is a severely under-converged $8\times 8\times 8$ $\mathbf{k}$-point grid. The spectrum produced by this coarse grid alone shows numerous spurious peaks (see Fig. \ref{fig:1silicon}), which reveals a high degree of artificial localisation of the excitons imposed by the $8\times 8\times 8$ $\mathbf{k}$-grid. We then took the latter as the coarse grid for the double grid method and added a fine grid of $\mathbf{\kappa}$-points to it. Fig. S1 shows that a fine (double) grid of $24\times 24\times 24$ $\mathbf{\kappa}$-points on top of this coarse grid immediately suppresses this artificial localisation. Denser double grids improve upon this result (See Fig. S2). Ultimately, the spectrum obtained with a $60\times 60\times 60$ fine $\mathbf{\kappa}$-grid on top of an $8\times 8\times 8$ coarse $\mathbf{K}$-grid is in close agreement with experiments (see Fig. \ref{fig:1silicon}). The comparison here is done with experimental data at 10 K available in the literature for Si bulk~\cite{Jellison1983}. 

\subsection{GaAs bulk}
\begin{figure}[h]
    \centering
    \includegraphics{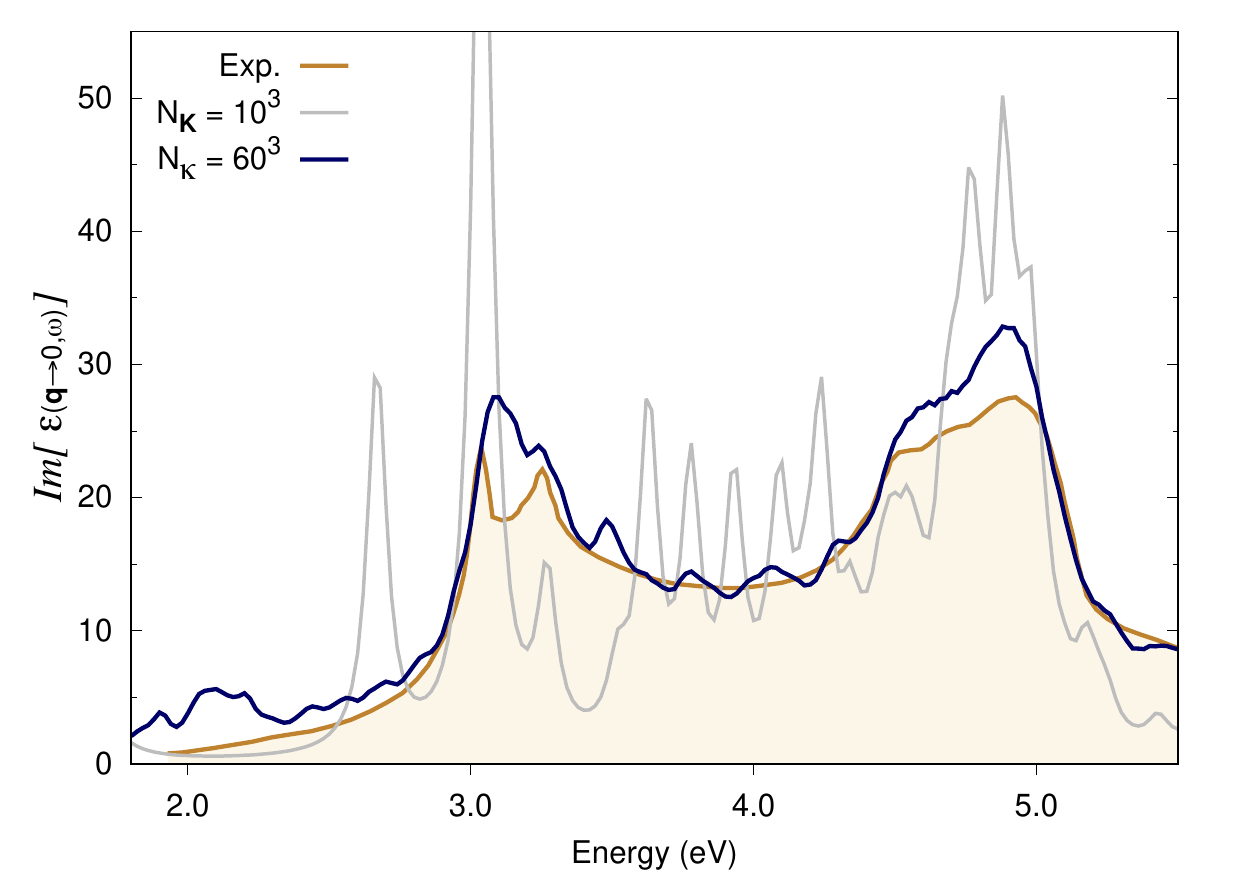}
    \caption{Optical absorption spectra of bulk GaAs. BSE spectra calculated using the double-grid approach described in this work are assessed against the experimental spectrum at 22 K from Ref.\cite{Festkorperforschung1987}. Spectra are calculated on an $10\times 10\times 10$ coarse $\mathbf{k}$-grid and a denser fine $\mathbf{k}$-grid, indicated by $N_\mathbf{\kappa}$. $N_\mathbf{K}=10^3$ corresponds to a standard calculation on an $10\times 10\times 10$ $\mathbf{k}$-grid while $N_\mathbf{\kappa}=60^3$ corresponds to a double-grid calculation with a $60\times 60 \times 60$ fine grid. We consider all $e-h$ pairs from the top 4 valence bands to the 4 bottom conduction bands. All $\mathbf{k}$-grids are Gamma-centred.}
    \label{fig:2GaAs}
\end{figure}

As in the case of Si, GaAs also requires very dense k-sampling for its optical response to be converged.  The coarse grid in this case is an under-converged $10\times10\times10$ Gamma-centred $\mathbf{k}$-point grid.  Indeed, the spectrum produced by this coarse grid alone also presents various spurious peaks, revealing a high degree of artificial localisation of its excitons (see Fig. \ref{fig:2GaAs}). Fig. S3 shows that adding a fine (double) $\mathbf{\kappa}$-grid of $20\times 20\times 20$ does not solve the problem fully. However, the spectra with $40\times 40\times 40$ or $60\times 60\times 60$ $\mathbf{\kappa}$-grids match the experimental data relatively well (see Fig. S4 and Fig. \ref{fig:2GaAs}, respectively). In particular, the latter grid appears to capture the splitting of the first exciton present in the experimental data, although a denser coarse grid would be required as a better starting point to fully reproduce this feature. Indeed, this result still seems to suffer from slight artificial localisation, e.g., at 3.5~eV. The comparison is drawn to available experimental data for GaAs at 22~K~\cite{Festkorperforschung1987}.

\subsection{MoS$_2$ monolayer}
The convergence of the absorption spectrum of monolayer MoS$_2$ with the $\mathbf{k}$-grid within BSE has been discussed in the appendix of Ref. \cite{Molina-Sanchez2013}. The latter study reports a converged spectrum using a very dense $\mathbf{k}$-point grid and clarifies that when spin-orbit coupling is not considered, the first excitonic peak should not show any splitting. In fact, other works which used an under-converged $\mathbf{k}$-grid, mistakenly showed a splitting of the first excitonic peak in collinear spin-polarised calculations with no spin-orbit coupling accounted for. 
 
\begin{figure}[h]
    \centering
    \includegraphics{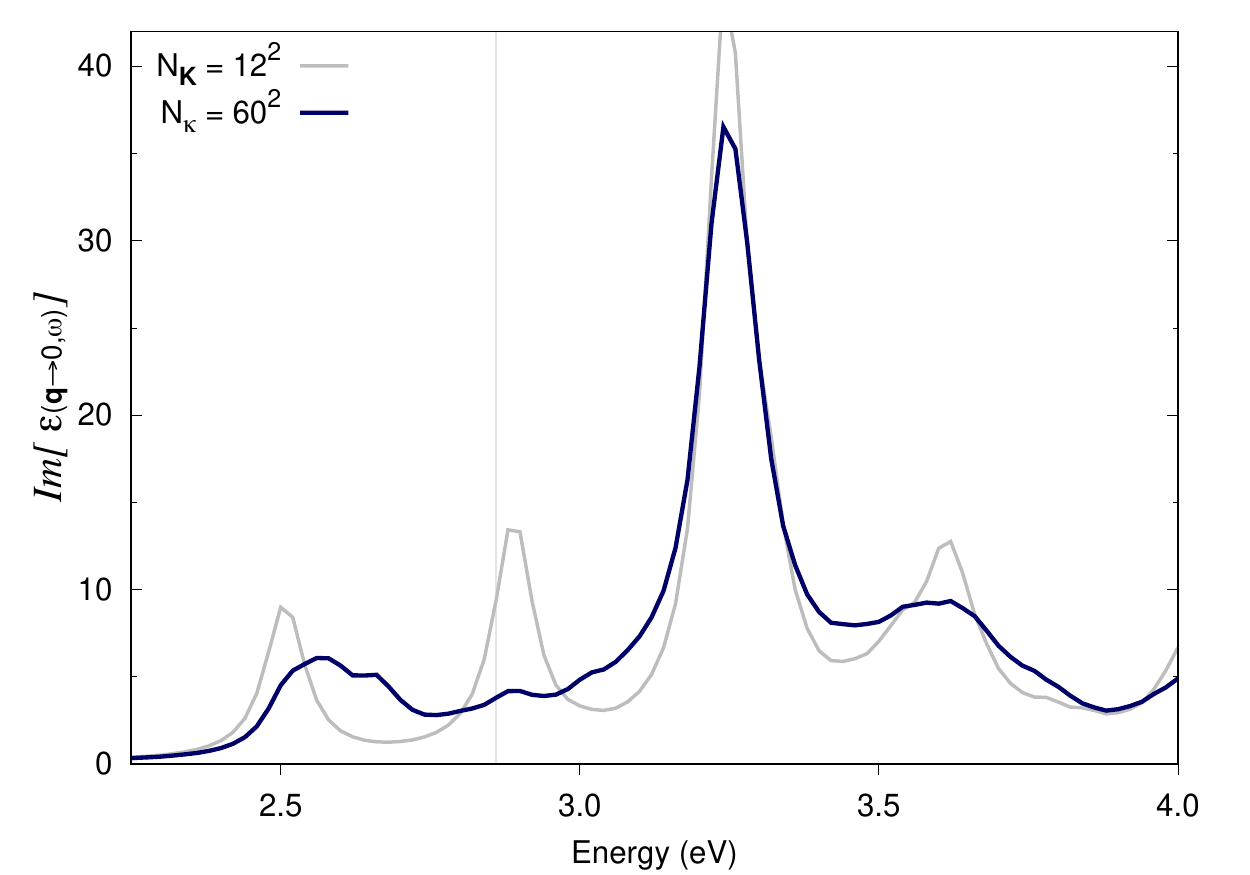}
    \caption{Optical absorption spectra of monolayer MoS$_2$. BSE spectra calculated using the double-grid approach described in this work. Spectra are calculated on an $12\times 12\times 1$ coarse $\mathbf{k}$-grid and a progressively denser fine $\mathbf{k}$-grid, indicated by $N_\mathbf{\kappa}$. $N_\mathbf{K}=12^2$ corresponds to a standard calculation on an $12\times 12\times 1$ $\mathbf{k}$-grid while $N_\mathbf{\kappa}=60^3$ corresponds to a double-grid calculation with a $60\times 60 \times 1$ fine grid. We consider all $e-h$ pairs from the top 3 valence bands to the 5 bottom conduction bands. All $\mathbf{k}$-grids are Gamma-centred.}
    \label{fig:4MoS2}
\end{figure}

Here, we aim to reproduce the spectra in Ref.~\cite{Molina-Sanchez2013} via the double-grid method (i.e., at a fraction of the computational cost). Our results follow a similar trend to the spectrum thereby reported. In the work of Molina-Sanchez \textit{et al.}, spectra with $12\times 12 \times 1$ or $18 \times 18\times 1$ (single grids) show splitting of the first peak while (single) $\mathbf{k}$-grids of $24\times 24\times 1$ or $30\times 30\times 1$ solve the issue. Our result with a $12\times 12\times 1$ single grid is equivalent to that of Ref. \cite{Molina-Sanchez2013} (except for a rigid shift in energy), i.e., it shows undue splitting (see Fig. \ref{fig:4MoS2}). Adding a double grid of $24\times 24\times 1$ $\mathbf{\kappa}$-points also results in undue splitting while $48\times 48\times 1$ appears to eliminate it (see Figs. S5 and S6, respectively). A double grid of $60\times 60\times 1$ further improves upon this result (see Fig. \ref{fig:4MoS2}). Importantly, the quality of the double grid spectrum with a double grid of $60 \times 60\times 1$ $\mathbf{\kappa}$-points is not far from what was achieved in Ref. \cite{Molina-Sanchez2013} with a $30\times 30\times 1$ single grid. The double grid approach correctly captures the physics at play despite representing roughly the computational cost of a $12\times 12\times 1$ regular BSE Haydock calculation (see Sec.~\ref{sc:discussion}).

\section{Discussion}\label{sc:discussion}
\subsection{Computational cost}
As described above, Lanczos approaches to the BSE eliminate the need to invert the BSE kernel or fully diagonalise the two-particle Hamiltonian, which would become the bottleneck of the calculation whenever required. Instead, Lanczos solvers replace these highly demanding tasks by very efficient and computationally inexpensive iterative schemes. This numerical advantage means that the solution step itself does not drive the computational cost any longer, but rather, computing and storing the BSE matrix now becomes the bottleneck of the calculation. The method proposed in this work addresses this issue directly. First, the KS orbitals in the fine grid need not be available, i.e., not stored nor loaded into memory. Moreover, the kernel matrix elements in the fine grid, and consequently, the corresponding oscillator strengths, need not be calculated. As a result, the size of the BSE kernel matrix will effectively be that of the coarse grid. For instance, if we consider a coarse grid of $10\times 10\times 10$ and a fine grid of $60\times 60\times 60$, then there would be 1000 $\mathbf{K}$-points and 216000 $\mathbf{\kappa}$-points. 
The full BSE kernel would have $(200 \times N_c \times N_b)^2$ more matrix elements than the approximated one. Depending on the number of bands required for convergence, the steps of computing and storing that many matrix elements may draw the line between what is feasible and what is not, not only in terms of processing power, but also due to memory and disk-storage limitations.

\begin{figure}
    \centering
    \includegraphics[scale=1.4]{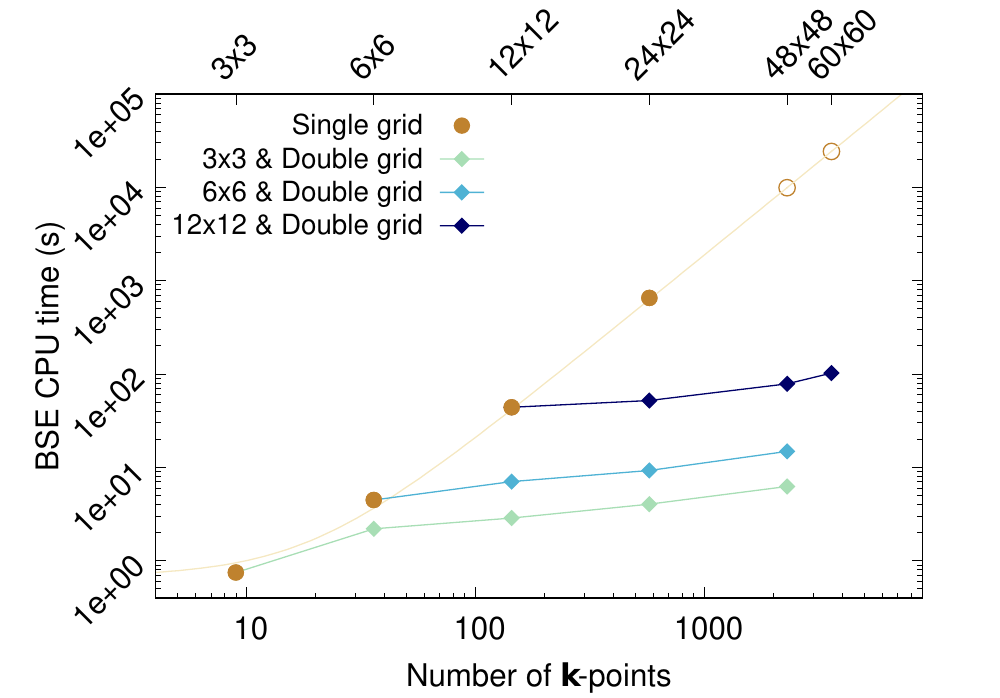}
    \caption{CPU time in seconds required to calculate and store the BSE kernel, and solve it via Haydock's iterative scheme as a function of the number of $\mathbf{k}$-points. Brown circles represent \textit{full} BSE calculations. The data for $48\times 48\times 1$ and $60\times 60\times 1$ $\mathbf{k}$-grids have been estimated via a quadratic fitting. The diamonds denote double-grid BSE calculations. Lines connect double-grid calculations using the same coarse-grid. All calculations were carried out in one processor for comparability purposes. We consider all $e-h$ pairs from the top 3 valence bands to the 5 bottom conduction bands.}
    \label{fig:7computational_cost}
\end{figure}

Let us consider monolayer MoS$_2$ to address how the computational cost compares between our double grid approach with a given fine $\mathbf{\kappa}$-grid and the regular (\textit{full}) BSE calculation using that same fine grid as the only (single) $\mathbf{k}$-grid. 
Fig.~\ref{fig:7computational_cost} shows the combined time required to calculate the BSE kernel and solve the eigen-problem via Haydock's scheme as a function of $\mathbf{k}$-points used in the calculations. For comparability purposes, all the calculations shown in Fig. \ref{fig:7computational_cost} have been carried out with just one processor. 
For the \textit{full} solution of the problem (brown circles) the number of $\mathbf{k}$-points has quadratic scaling from one $\mathbf{k}$-grid to another (as it does for any 2D material) and the CPU time scales quadratically with the total number of $\mathbf{k}$-points. This latter dependence stems from the size of the e-h basis set and the number of matrix elements of the BSE kernel, i.e., $(N_\mathbf{K} \times N_c \times N_b)^2$. 
The computational cost of the double grid approach proposed in this work (green-blue diamonds connected by lines) increases only slightly with the size of the fine-grid, when the same coarse grid is used. Since the BSE kernel is calculated only in the coarse-grid, this increase is due to the Haydock solver, 
which now has to process larger Haydock vectors. Nonetheless, it is apparent that the increased CPU time due to Haydock is minor and far more manageable than the scaling of the \textit{full} BSE problem. 
Overall, fine grid has little impact on the CPU time required by the method we propose. In fact, Fig. \ref{fig:7computational_cost} clearly shows that the computational cost of the double grid method is roughly driven by the coarse grid.

\subsection{Kernel extension to fine grid}\label{sc:an_DKE}

The kernel extension is the key approximation used in the double-grid approach. The DKE in Sec.~\ref{sc:methods} was selected for its simplicity and low computational cost. There are other possible kernel extensions. In particular, we also considered a kernel extension with similar characteristics, that we refer to as full kernel extension (FKE). First, let us define the FKE approach formally: FKE implies that each matrix element of the coarse grid BSE kernel is expanded into an all-ones block times the original matrix element, which leads to

\begin{equation}
    \Xi_{\substack{ nm\kappa_{\mathbf{I}_i}\\n'm'\kappa_{\mathbf{I'}_{i'}}}} \equiv \;
    \Xi_{\substack{ nm\mathbf{K_I}\\n'm'\mathbf{K_{I'}}}} \forall  i,i'
    \label{eq:ec_FKE}
\end{equation}
(see SI for a visual representation of Eq. \ref{eq:ec_FKE}). As a result, the way in which the fine-grid matrix vector multiplication is carried out also differs from DKE. In FKE, this operation is performed as 

\begin{equation}
    r_{nm\kappa_{\mathbf{I}_{i}}} =
    \sum_{n'm'\mathbf{K}_{\mathbf{I'}}} 
    \Xi_{\substack{ nm\mathbf{K}_{\mathbf{I}}\\n'm'\mathbf{K}_{\mathbf{I'}}}}
    \sum_{\substack{i' \; \in \\ \mathbf{Dom}(\mathbf{K_{I'}})}}
    c_{n'm'\kappa_{\mathbf{I'}_{i'}}}. 
    \label{eq:ec_FKE_KbyV2}
\end{equation}
Eqs. \ref{eq:ec_FKE} and \ref{eq:ec_FKE_KbyV2} of FKE are analogous to Eqs. \ref{eq:ec_DKE} and \ref{eq:ec_KbyV2} of DKE, respectively. 

\begin{figure}[h]
    \centering
    \includegraphics{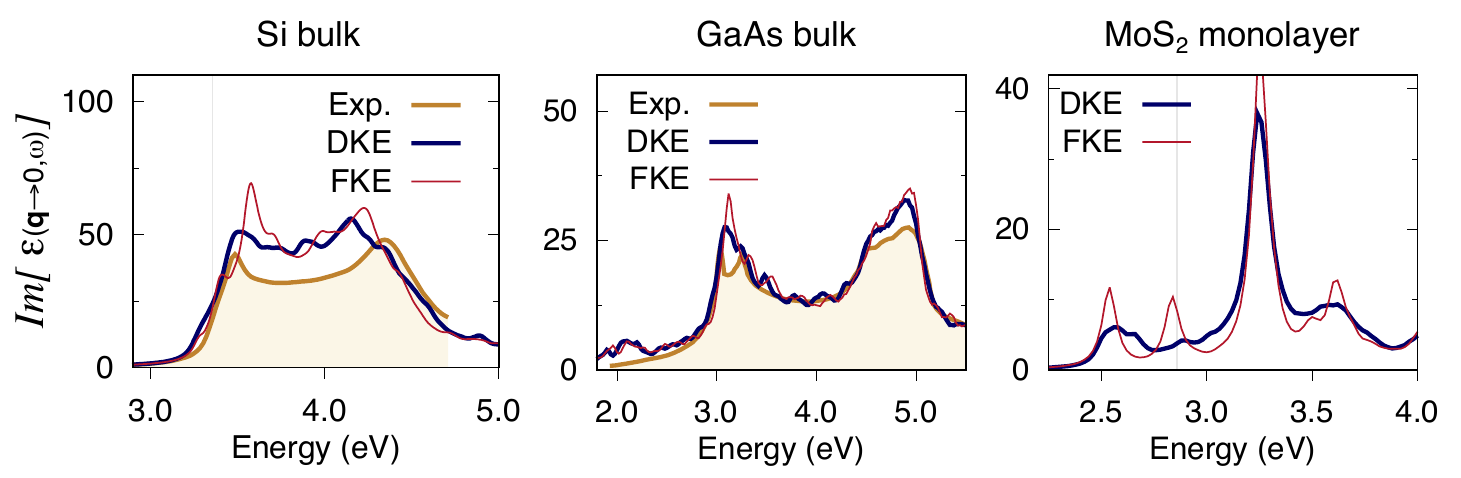}
    \caption{Optical absorption spectra of bulk Si (left panel), GaAs (mid panel) and MoS$_2$ monolayer (right panel).  Comparison of spectra obtained by diagonal kernel extension (DKE) and full kernel extension (FKE). For Si and GaAs the coarse $\mathbf{k}$-grid was $8\times 8\times 8$ and $10\times 10\times 10$ respectively and the fine $\mathbf{k}$-grid, $60\times 60\times 60$. For the MoS$_2$ monolayer, the coarse $\mathbf{k}$-grid was $12\times 12\times 1$ and the fine $\mathbf{k}$-grid, $60\times 60\times 1$.}
    \label{fig:5DKE_FKE}
\end{figure}

In terms of the spectra produced by either kernel extensions, the comparison consistently favours DKE over FKE in all the materials tested in this work. The difference may be less noticeable in systems with weaker excitonic effects. A comparison for the materials in Sec.~\ref{sc:results} is shown in Fig.~\ref{fig:5DKE_FKE}. In the case of Silicon (Fig. \ref{fig:5DKE_FKE}A), it is apparent that DKE is better than FKE at suppressing the artificial localisation found around 3.6 eV. For GaAs (Fig. \ref{fig:5DKE_FKE}B), DKE also shows an improvement with respect to FKE when dealing with the artificial localisation at around 3.1 eV. Finally, monolayer MoS$_2$ (Fig. \ref{fig:5DKE_FKE}C) also follows the trend found in this work, i.e., DKE is consistently better than FKE. In this case in particular, the difference between both approaches is very stark. In fact, the FKE approach shows little to none improvement with respect to the $12\times 12\times 1$ single $\mathbf{k}$-grid as far as the first exciton is concerned (cf. Fig. \ref{fig:4MoS2}).

\begin{figure}[p]
    \centering
    \includegraphics{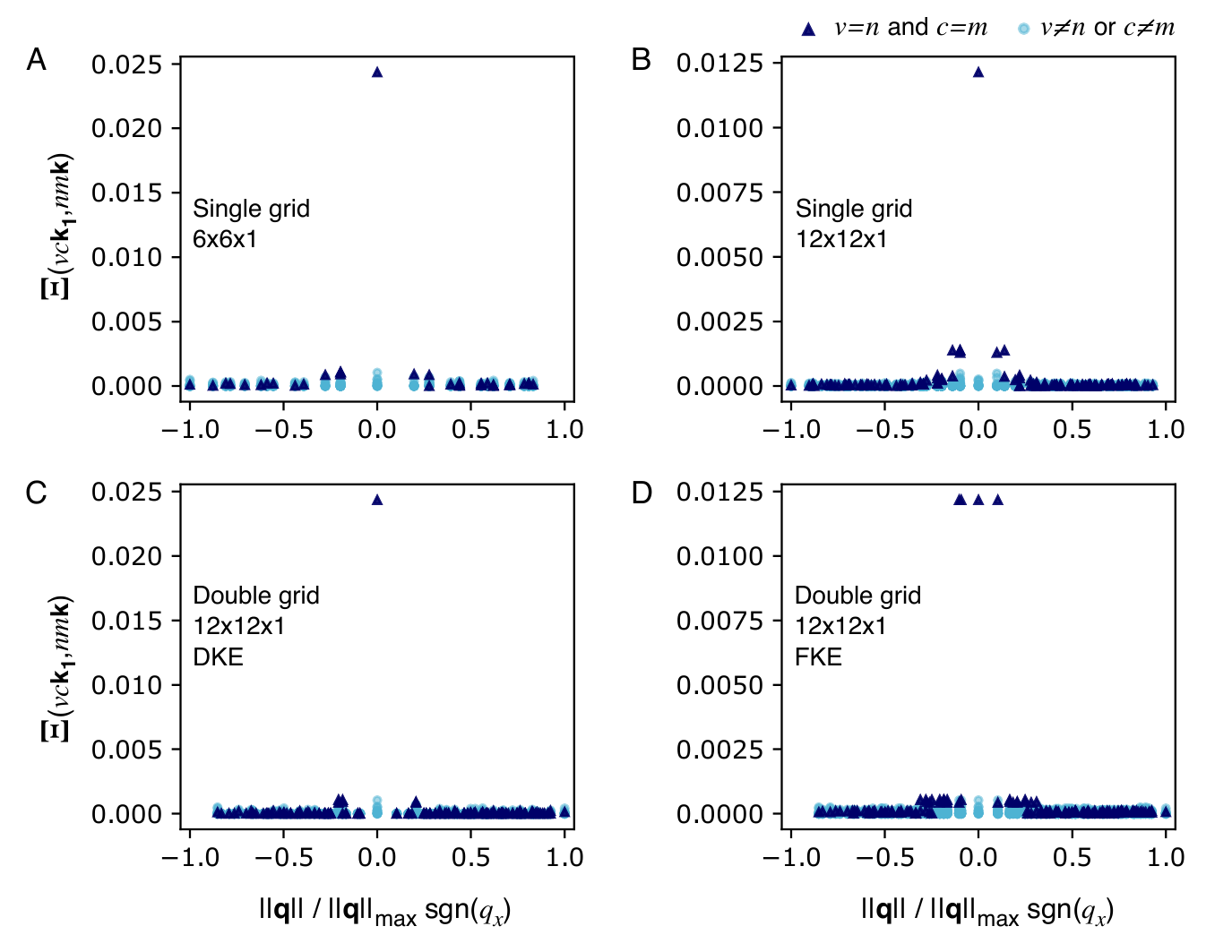}
    \caption{Module of the BSE Kernel matrix element between one transition ($vc\mathbf{k}_1$) and every other transition in the e-h space ($nm\mathbf{k}$). The data plotted here corresponds to MoS$_2$ with all the bands required for convergence. Panel A shows the matrix elements considering only a single grid of $6\times6\times1$ $\mathbf{k}$-points. The DKE and FKE matrix elements are obtained from a $6\times6\times1$ coarse $\mathbf{K}$-grid and a $12\times12\times1$ double $\mathbf{\kappa}$-grid. The fine grid data is simply what DKE and FKE try to approximate, i.e., the kernel matrix elements obtained with one single grid of $12\times12\times1$ $\mathbf{k}$-points.}
    \label{fig:6c-K_q}
\end{figure}

In order to explain the better performance of DKE over FKE, we will discuss the properties of the BSE kernel and the two-particle Hamiltonian matrices, which are related by Eq.~\ref{eq:ec_2pH}. In general, the kernel matrix elements $\Xi_{\substack{ nm\mathbf{k}\\n'm'\mathbf{k'}}}$ are sharply peaked at $\mathbf{q}=0$ \cite{Rohlfing1998,Rohlfing2000}, i.e., for $\mathbf{k}=\mathbf{k'}$.
This does not mean that every matrix element with $\mathbf{q}=0$ will have a higher value than the remaining matrix elements. In fact, that is only true for the diagonal elements $\Xi_{\substack{ nm\mathbf{k}\\nm\mathbf{k}}}$, while the $\mathbf{q}=0$ elements coupling different sets of bands ($\Xi_{\substack{ nm\mathbf{k}\\n'm'\mathbf{k}}}$) are closer in value to all other $\mathbf{q} \ne 0$ matrix elements. This is again exemplified with monolayer MoS$_2$ in Fig.~\ref{fig:6c-K_q}. The latter shows the module of every matrix element between a given transition ($v=13$, $c=14$ and $\mathbf{k}_{1}=(-0.166,-0.166,0)$) and every other transition in the e-h space, i.e., one row of the BSE kernel matrix. This data is plotted as a function of the magnitude $||\mathbf{q}||/||\mathbf{q}||_{max} \; sgn(q_x)$, where $\mathbf{q} = \mathbf{k} - \mathbf{k_1}$. Fig.~\ref{fig:6c-K_q}A shows the BSE kernel as obtained with a single grid of $6\times 6\times 1$ $\mathbf{k}$-points, where we can see that the diagonal matrix element (the selected transition with itself) is an order of magnitude higher than all other matrix elements (many of which also have $\mathbf{q}=0$). The fine grid of $12\times 12\times 1$ $\mathbf{k}$-points better captures the build-up to the peak of the graph as it has many more  $\mathbf{k}$-points around the selected one (see Fig. \ref{fig:6c-K_q}B). Unfortunately, the double grid approach proposed here cannot capture this feature because it is meant not to imply any extra computation or storage of matrix elements at fine grid $\mathbf{\kappa}$-points. However, the reader should bear in mind that while this feature is missing in our approximated BSE kernel, the benefits of this double grid approach reside in exactly knowing the transition energies at the fine grid $\mathbf{\kappa}$-points (see SI for detailed discussion). At this point, what we expect from the approximated kernel is not to introduce unphysical matrix elements, and in this regard DKE performs much better than FKE. Fig. \ref{fig:6c-K_q}C shows how the BSE kernel matrix elements approximated by DKE still represent a function of $\mathbf{q}$ that is sharply peaked at the origin. Conversely, the FKE approach means that many matrix elements in $\mathbf{Dom}(\mathbf{k}_{1})$, and consequently at $\mathbf{q\neq 0}$, will take the value of the peak. We know that such behaviour as a function of $\mathbf{q}$ would not arise should more $\mathbf{k}$-points be included (see \ref{fig:6c-K_q}B). Hence, we believe DKE constitutes a better approximation of the BSE kernel than FKE.
Further arguments in favour of the DKE over the FKE are presented in SI.

\subsection{Limitations of the approach}
The double-grid approach presented in Sec.~\ref{sc:dgrid} is based on two approximations: the DKE (Eq. \ref{eq:ec_DKE}) and the approximation of the starting Haydock vector (Eq. \ref{eq:ec_initialisation_FG}).
The DKE has been extensively analysed in Sec.~\ref{sc:an_DKE}. From the analysis, it emerges that the predominance of the matrix elements with $\mathbf{q} \approx 0$ is crucial to the success of the approximation. This is consistent with the spatial de-localisation of the exciton over many unit cells. 
Conversely, when the exciton is localised on few unit cells---as it is the case for instance in wide-gap insulators---the approximation may break down because of the significant contribution to the BSE kernel of matrix elements with $\mathbf{q} \neq 0$. We verified this is the case, for example, for bulk hexagonal boron nitride (h-BN).
The breakdown of the approach for these cases is, however, not critical. In fact, excitons that are localised on few unit cells can be described accurately with a modest $\mathbf{k}$-point sampling and the double-grid is not needed.

\begin{figure}[h]
    \centering
    \includegraphics{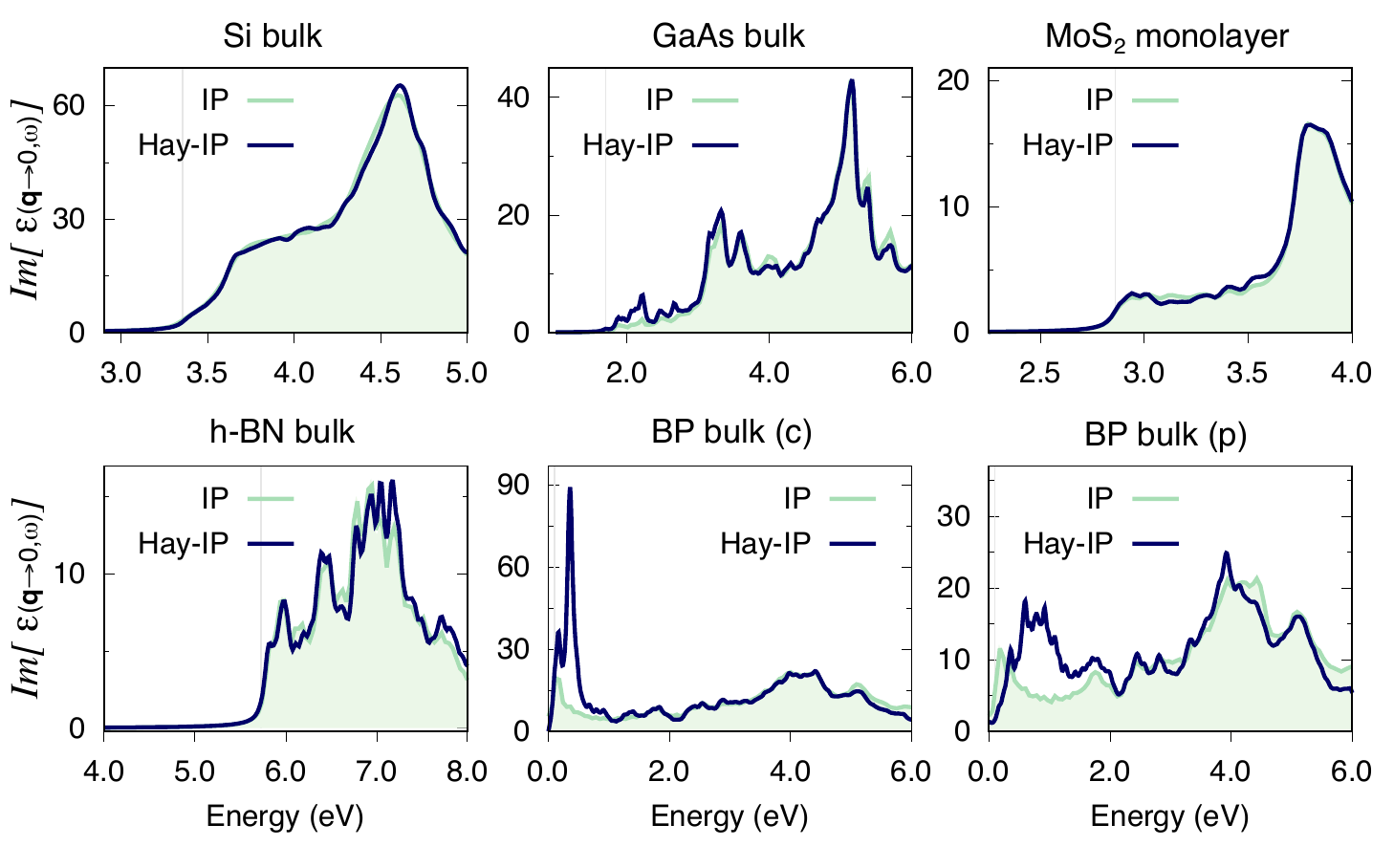}
    \caption{Optical absorption spectra at the IP level calculated with the double grid method (labelled Haydock-IP) and via a full calculation on a fine grid (labelled IP), for all materials considered in this study. In all cases, the fine grid in the double-grid Haydock-IP calculation is of the same dimensions as the fine grid used in full for the IP calculation. The coarse grids and double/fine grids used for each material are listed below. Si: $8\times8\times8$ and $60\times60\times60$, GaAs: $10\times10\times10$ and $40\times40\times40$, MoS$_2$: $12\times12\times1$ and $60\times60\times1$, h-BN: $12\times12\times4$ and $24\times24\times8$, BP(c): $14\times10\times4$ and $42\times30\times12$, BP(p): $5\times5\times6$ and $30\times30\times36$. All $\mathbf{k}$-grids are Gamma-centred.}
    \label{fig:8IP_HayIP}
\end{figure}

The approximation for the starting Haydock vector (Eq. \ref{eq:ec_initialisation_FG}) implies the assumption that (within the length gauge and dipole approximation) the dipole matrix elements in the fine grid can be approximated by those in the coarse grid, namely, 
\begin{align}
\langle n\kappa_{\mathbf{I}_i}|\;\hat{\mathbf{r}}\;|m\kappa_{\mathbf{I}_i} \rangle \approx \langle n\mathbf{K_I}|\;\hat{\mathbf{r}}\;|m\mathbf{K_I} \rangle,
\label{eq:approx_dip}
\end{align}
for $\kappa_{\mathbf{I}_i} \in \mathbf{Dom}(\mathbf{K_I})$, where $\hat{\mathbf{r}}$ is the position operator. This assumption can be verified at the level of the independent particle approximation (IPA) by comparing the IPA spectrum obtained with the double-grid approach (which we call Haydock-IP) with the IPA spectrum calculated on the fine grid. In fact, in the independent particle case, Eq.~\ref{eq:approx_dip} is the only approximation introduced by the double grid. For the systems considered in Sec.~\ref{sc:results}, we verified that indeed the IPA spectra obtained within the double-grid approach agree well with the IPA calculated on the corresponding fine grid (see Fig. \ref{fig:8IP_HayIP}). It is also interesting to note that this particular approximation is valid for h-BN, which singles out the BSE kernel ($\mathbf{q}\approx 0$) approximation as the only factor hindering the application of the double-grid method to this material. In particular, GaAs shows a minor discrepancy in the IPA spectra around 2.1~eV (see Fig. \ref{fig:8IP_HayIP}), a region of the spectrum where $\mathbf{k}$-point convergence is markedly difficult. This is due to the steep dispersion of the conduction band of GaAs around the Gamma point, where the optical gap occurs (see, for example, \cite{Festkorperforschung1987}). As a result, the approximation of the oscillator strengths around Gamma by the corresponding matrix element at Gamma (Eq.~\ref{eq:approx_dip}) is a rather poor one, which translates into an unphysical feature around 2.1~eV in the BSE spectrum as well (see Fig. \ref{fig:2GaAs}).

There are also instances in which the approximation in Eq. \ref{eq:approx_dip} breaks down substantially. As an example, Fig.~\ref{fig:8IP_HayIP} shows this breakdown for the optical absorption of bulk black-phosphorous (BP) along the armchair direction \cite{Tran2014}. The IPA spectrum obtained within the double-grid approach has strong peaks around 0.3~eV which are not present in the reference calculation. The appearance of this artefact can be understood considering that the dipole matrix elements (Eq.~\ref{eq:approx_dip}) are calculated as
$\frac{\langle n\mathbf{K}_I|\;\hat{\mathbf{v}}\;|m\mathbf{K}_I \rangle}{E_{nm\mathbf{K}_I}}$, where $\hat{\mathbf{v}}$ is the velocity operator.
BP has a minimum KS band-gap of about 0.2~eV (0.1~eV at the DFT level) and thus the corresponding dipole matrix element is large. Within the double-grid approach, all fine-grid $\mathbf{k}$-points in the domain of the $\mathbf{k}$-point corresponding to the minimum KS band-gap use the same value which largely overestimates the actual dipole matrix element. Notably, carrying out the calculations in the primitive rather than in the conventional unit cell (see Fig.~\ref{fig:8IP_HayIP}), improves the agreement with the reference IPA fine-grid spectrum, suggesting that in this case the coarse grid does a better job at sampling the Brillouin zone around the $\mathbf{k}$-point corresponding to the minimum KS band-gap. Nevertheless, this Haydock-IP spectrum still presents artificial features between 0.5--1.0~eV, preventing the application of the double-grid method presented in this work to BP.

Further to note is that using Eq.~\ref{eq:approx_dip} the extension from the coarse to the full grid is done for the position dipoles, i.e. within the length gauge. This implies that, to ensure gauge invariance~\cite{PhysRevB.95.155203},   
$\langle n\kappa_{\mathbf{I}_i} |\; \hat{\mathbf{v}} \;| m\kappa_{\mathbf{I}_i} \rangle =\langle n\mathbf{K_I} |\; \hat{\mathbf{v}} \;| m\mathbf{K_I} \rangle (E_{nm \kappa_{\mathbf{I}_i}} / E_{nm \mathbf{K_I}})$.    
An alternative choice could be instead to assume 
$\langle n\kappa_{\mathbf{I}_i} |\; \hat{\mathbf{v}} \;| m\kappa_{\mathbf{I}_i}\rangle =\langle n\mathbf{K}_I |\; \hat{\mathbf{v}} \;| m\mathbf{K}_I \rangle$,
i.e. to perform the extension of the velocity matrix elements and  accordingly obtain
$\langle n\kappa_{\mathbf{I}_i} |\; \hat{\mathbf{r}} \;| m\kappa_{\mathbf{I}_i} \rangle =\langle n\mathbf{K}_I |\; \hat{\mathbf{r}} \;| m\mathbf{K}_I \rangle (E_{nm \mathbf{K}_I} / E_{nm \kappa_{\mathbf{I}_i}}).$    
Preliminary results show that such choice may in fact lead to better results for BP.

\subsection{Workflow implementation}
Based on this discussion, we can propose a workflow to assess whether a given material satisfies these approximations and could thus be described well by the double-grid method presented here. Firstly, a \textit{full} fine grid IP calculation must be converged with respect to $\mathbf{k}$-points. Alternatively one could choose the densest $\mathbf{k}$-grid that can be treated at the IP level, where limitations usually reside on memory. Let us take an example where this first step of the procedure returns a $60\times60\times60 \; \mathbf{k}$-grid. The next stage would be to find an appropriate coarse $\mathbf{k}$-grid, i.e., one that satisfies the approximation in Eq.~\ref{eq:approx_dip}. This would entail running several double-grid calculations at the IPA level (Haydock-IP) with varying coarse grid and with a fine grid of $60\times60\times60$. By matching the Haydock-IP spectrum to the \textit{full} $60\times60\times60$ fine-grid IP spectrum, a sound coarse grid can be chosen by means of a fairly inexpensive procedure (e.g., $8\times8\times8$). The next step in the workflow would necessarily involve BSE calculations as the approximation on the BSE kernel cannot be tested at the IPA level. With the chosen coarse grid, successive double-grid BSE calculations with varying fine grids should be carried out in order to converge the dimensions of the latter. It is worth mentioning that all these calculations have roughly the same computational cost and requirements, at the level of the coarse-grid $8\times8\times8$ BSE calculation. It should be highlighted that convergence of the fine-grid in the double-grid method does not guarantee the validity of the DKE approximation. At this point, one should turn to available data, either experimental or theoretical, in order to assess the validity of the results on physical grounds.

\section{Conclusions}
In this work, we presented a double grid approach to the problem of $\mathbf{k}$-point sampling in the solution of the BSE equation for the calculation of optical spectra of semiconductors. This responds to the fact that very dense $\mathbf{k}$-point grids are required for BSE calculations to be fully converged due to the large periodicity of excitonic wavefunctions, usually reaching several supercells. This sampling requirement is the bottleneck in BSE calculations and, for a wide variety of solids, this imposes a computational burden that renders the calculation prohibitively costly. We tackled this challenge by applying a double grid approach to the computationally cheapest among the BSE solvers, i.e., the Lanczos-based Haydock scheme, thus maximising the size and range of materials for which this method could be useful. Our double grid approach is based on combining a coarse $\mathbf{k}$-grid where both KS eigen-values and eigen-vectors are known with a fine $\mathbf{k}$-grid where only KS energies are required, which eases memory and disk storage requirements. With this strategy, the coarse $\mathbf{k}$-grid drives the computational cost while the $\mathbf{k}$-fine grid tries to capture the physics of spread out excitons in an approximated way without requiring significant extra computation. 

This scheme was implemented in the Yambo code (see SI for availability) and tested for bulk Si, bulk GaAs and monolayer MoS$_2$, all of which are known to require very dense $\mathbf{k}$-point grids to achieve convergence. The results are satisfactory in all cases, reproducing data reported elsewhere with a relatively low computational cost close to that of the coarse grid alone. There is a slight increase in the CPU time required by the Haydock step, however this scales very favourably with increasingly dense $\mathbf{k}$-meshes, at variance with regular non-double grid approaches. Different ways to extend the BSE kernel calculated in the coarse grid to the fine grid are discussed and compared, determining that the so-called diagonal kernel extension is the preferred method. 

The approximations introduced with the double-grid approach have been discussed, together with the limits they impose to its validity. On the one hand, the diagonal kernel extension limits the applicability of this approach to systems with excitons delocalised over many unit cells. On the other hand, the latter are precisely the main target of the double-grid approach, given that spatially localised excitons are usually well described by a relatively coarse $\mathbf{k}$-grids. Further, we discussed how the validity of the approximation on the dipole matrix elements can be verified and controlled with inexpensive calculations at the level of the independent particle approximation.

In light of the promising results achieved by the double grid approach presented in this work, considering its simplicity and taking into account its compatibility with the very efficient Lanczos based BSE solution schemes (i.e., Haydock), we hope our work will facilitate the calculation of optical spectra in semiconductors that could not be computationally afforded to date.

\section*{Conflict of Interest Statement}
The authors declare that the research was conducted in the absence of any commercial or financial relationships that could be construed as a potential conflict of interest.

\section*{Author Contributions}
IMA contributed to the development of the approach and  implemented it into Yambo. IMA also carried out the calculations, analysed their results and wrote most of the manuscript. DS put forward the idea of the approach---that was then improved together with IMA and MG---assisted with the implementation of the approach and contributed to the analysis of the results. MG contributed to the development of the approach and assisted with its implementation. MG also contributed to the analysis of the results and wrote a part of the manuscript.

\section*{Funding}
MG is grateful for support from the Engineering and Physical Sciences Research Council, under grant EP/V029908/1.  
DS acknowledges support from Italian Ministry of Education, University and Research (MIUR) through the Research Project of National Relevance (PRIN) BIOX under grant No. 20173B72NB; from the European Union project MaX (Materials design at the eXascale) H2020-EINFRA-2015-1 (Grant Agreement 824143) and from the Nanoscience Foundries and Fine Analysis-Europe H2020-INFRAIA-2014-2015 (Grant Agreement No. 654360). 

\section*{Acknowledgments}
The authors acknowledge useful discussions with Daniele Varsano, Matteo Zanfrognini and Claudio Attaccalite. 

\bibliographystyle{naturemag} 
\bibliography{Bibliography.bib}

\end{document}